\documentclass[12pt]{article}
\pdfoutput=1
\usepackage{amsmath,amssymb,graphicx, float} 
\usepackage{epsf}
\usepackage{pstricks}
\usepackage[sort&compress,numbers]{natbib}
\usepackage{ctable}
\usepackage{multirow}

\newcolumntype{C}[1]{>{\centering\let\newline\\\arraybackslash\hspace{0pt}}m{#1}}

\newcommand{\beq}{\begin{eqnarray}}
\newcommand{\eeq}{\end{eqnarray}}

\newcommand\ZZ{\hbox{\zfont Z\kern-.4emZ}}
\font\zfont = cmss10 

\newcommand{\met}{\mbox{${\rm \not\! E}_{\rm T}$}}

\def\gappeq{\mathrel{ \rlap{\raise.5ex\hbox{$>$}}
                      {\lower.5ex\hbox{$\sim$}}  } }
\def\lappeq{\mathrel{ \rlap{\raise.5ex\hbox{$<$}}
                      {\lower.5ex\hbox{$\sim$}}  } }
\begin{document}
\begin{titlepage}

\vskip.5cm
\begin{center}
{\huge \bf General Neutralino NLSPs}

\bigskip

{\huge \bf at the Early LHC}

\vskip.1cm
\end{center}
\vskip0.2cm

\begin{center}
{\bf Joshua T.~Ruderman$^a$ and David Shih$^{b}$}
\end{center}
\vskip 8pt

\begin{center}
{\it $^a$Department of Physics\\
Princeton University, Princeton, NJ 08544}\\
\vspace*{0.2cm}
{\it $^b$ New High Energy Theory Center \\
Department of Physics and Astronomy \\
Rutgers University, Piscataway, NJ 08854}\\
\vspace*{0.1cm}

\end{center}

\vglue 0.3truecm

\begin{abstract}
\vskip 5pt \noindent Ê

Gauge mediated supersymmetry breaking (GMSB) is a theoretically well-motivated framework with rich and varied collider phenomenology. In this paper, we study the Tevatron limits and LHC discovery potential for a wide class of GMSB scenarios in which the next-to-lightest superpartner (NLSP) is a promptly-decaying neutralino. These scenarios give rise to signatures involving hard photons, $W$'s, $Z$'s, jets and/or higgses, plus missing energy. In order to characterize these signatures, we define a small number of minimal spectra, in the context of General Gauge Mediation, which are parameterized by the mass of the NLSP and the gluino. Using these minimal spectra, we determine the most promising discovery channels for general neutralino NLSPs. We find that the 2010 dataset can already cover new ground with strong production for all NLSP types. With the upcoming 2011-2012 dataset, we find that the LHC will also have sensitivity to direct electroweak production of neutralino NLSPs.

\end{abstract}

\end{titlepage}

\newpage

\renewcommand{\thefootnote}{(\arabic{footnote})}


\section{Introduction and Conclusions}\label{sec:intro}
\setcounter{equation}{0} \setcounter{footnote}{0}

The discovery era at the LHC is underway. Within the next two years, a large dataset is expected to be collected at 7 TeV\@. This is guaranteed to result in a rapid and dramatic expansion of the sensitivity to new physics, especially involving light colored states. Hopefully, we will even discover something.

As experimentalists plan their searches for new physics, now is the time to think carefully about how to design these searches optimally in order to cover as many scenarios as possible. It is also the time to think about how to report these results in the most meaningful way.

Gauge mediation is an extremely well-motivated and theoretically-sound supersymmetric scenario (for a review with original references, see~\cite{Giudice:1998bp}).  It automatically solves the SUSY flavor problem, and it provides a calculable and predictive framework. It also has rich and distinctive phenomenology~\cite{Dimopoulos:1996vz, Dimopoulos:1996fj, Dimopoulos:1996yq, Ambrosanio:1997rv, Culbertson:2000am, MatchevThomas, BMTW}. 
Recently, a model independent framework for gauge mediation was formulated in~\cite{Meade:2008wd, Buican:2008ws}. This has greatly expanded the possible parameter space for gauge mediation, and has led to a renewed interest in more general signatures~\cite{Meade:2009qv, Meade:2010ji, Ruderman:2010kj, Carpenter:2008he, Rajaraman:2009ga, Katz:2009qx, Abel:2009ve, Katz:2010xg, Abel:2010vb, Thalapillil:2010ek, Jaeckel:2011ma, Jaeckel:2011qj}.

In gauge mediation, the LSP is a nearly-massless gravitino. The lightest MSSM sparticle is then the next-to-LSP (NLSP), and it always decays in a universal way to the gravitino plus its SM superpartner. Since this decay rate is heavily suppressed by the SUSY-breaking scale, NLSP decays can be prompt or displaced; in this paper we will focus on the prompt case. This suppression of the decay rate also means that all heavier sparticles decay first down to the NLSP before decaying to the gravitino. Therefore, the nature of the NLSP is the most important aspect of the GMSB spectrum for collider signatures.

The most well-studied and well-searched-for gauge mediation scenario is where the NLSP is a bino-like neutralino. Then the discovery signature consists of $\gamma\gamma+\met$.  The LHC reach in this channel has been studied by several authors, including~\cite{Baer:1998ve, Nakamura:2010faa}. 

\begin{figure}[t!]
\begin{center}
\includegraphics[scale=0.6]{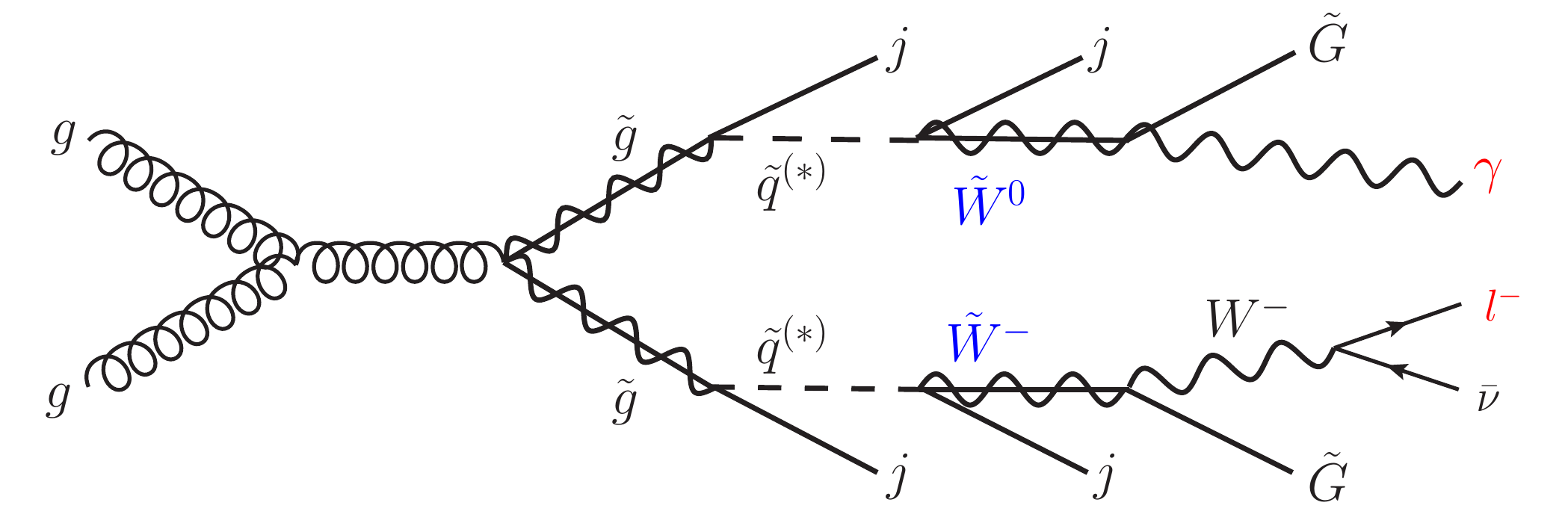}
\end{center}
\caption{One example process with colored production of wino co-NLSPs.  Here, two gluinos are pair produced, and each one decays through an on (or off) shell squark to two jets and a charged or neutral wino.  In this example, the neutral wino decays to a photon and a gravitino, while the charged wino decays to a leptonic $W$ and a gravitino.  The gravitinos carry missing energy, so that the signature is $l+\gamma+\mathrm{jets}+\met$.}
\label{fig:feynman}
\end{figure}

In this paper, we will study the simplest generalization of this, namely general neutralino NLSPs. Even this simple generalization leads to very interesting, much less well-explored signatures. These involve high $p_T$ leptons, $Z$'s, $W$'s, jets, and/or higgses plus missing energy. An example of such a signature is shown in figure~\ref{fig:feynman}. These signatures have been recently studied for the Tevatron in \cite{Meade:2009qv}. As discussed there, the phenomenology of general neutralino NLSPs is best understood by going to simplifying gauge eigenstate limits. Then the types are: bino NLSP, wino co-NLSP and higgsino NLSP\@. Higgsino NLSPs in turn can be classified by their decay modes -- they can decay dominantly to $Z$'s (Z-rich), to higgses (h-rich), or roughly 50/50 to $Z$'s and higgses (mixed).

The focus in \cite{Meade:2009qv} was on direct electroweak NLSP production at the Tevatron. We expand upon this analysis by including the possibility of strong production, and by exploring the LHC reach. In doing so, we will see explicitly the complementarity of the early LHC and the Tevatron with respect  to strong vs.\ electroweak production.

In traditional models of gauge mediation, such as Minimal Gauge Mediation \cite{Dine:1993yw,Dine:1994vc,Dine:1995ag}, the spectrum is controlled by a small number of parameters (and just one mass scale), and as a result, there are many relations among the sparticle masses. In particular, the colored sparticles (squarks, gluinos) are generally fixed to be much heavier than the electroweak sparticles. Therefore, the discovery potential at the early LHC, given limits from LEP and the Tevatron on the electroweak sparticles, is considerably reduced. 

In more general models of gauge mediation, however, there is no reason for the colored sparticles to be heavy. As was shown in \cite{Meade:2008wd,Buican:2008ws}, in the broader parameter space of General Gauge Mediation, there is no a priori relation between the colored and electroweak sparticle masses. All that must be satisfied are two simple sum rules for the squark masses. Therefore, the allowed colored sparticle masses in General Gauge Mediation can go much lower than in Minimal Gauge Mediation, and the potential for discovery at the early LHC is much greater.  

With these considerations in mind, we will formulate minimal spectra in this paper which allow for significant strong SUSY production at the early LHC, and populate the many final states available to general neutralino NLSPs. Our parameter spaces consist essentially of a colored production scale (gluino mass for simplicity) and the NLSP mass. We will decouple all other sparticles from the spectrum which are not essential for production or for the signature of interest. Our approach is very similar to the ``simplified models" approach utilized by many authors in recent phenomenological studies \cite{Dube:2008kf, Alwall:2008va, Alwall:2008ag, Alves:2011sq, SimplifiedModelWebsite,Meade:2009qv,Meade:2010ji,Ruderman:2010kj}.

\begin{table}[b]
\begin{center} \begin{tabular}{|c|c|c|c|c|c|}
\hline
channel & search & ref. & bino & wino & Z-rich higgsino \\
\hline
$ \gamma \gamma + \met$ & D\O~6.3 fb$^{-1}$  & \cite{D0ggmet} & \checkmark & \checkmark & \\
$\ell \gamma+\met$, & CDF 0.93 fb$^{-1}$ & \cite{cdflgX} & & \checkmark &  \\
jets + $\met$ & D\O~2.1 fb$^{-1}$&   \cite{D0jetsmet} & & \checkmark &\checkmark \\
$Z(e^+ e^-)$ + jets +$\met$ & CDF 2.7 fb$^{-1}$ &\cite{CDFWZmet} & & & \checkmark\\
\hline
\end{tabular} \end{center}
\caption{The strongest Tevatron limits on bino, wino and Z-rich higgsino NLSPs.}
\label{tab:TeVchannel}
\end{table}

\begin{table}[!b]
\begin{center} \begin{tabular}{|c|c|c|c|c|c|}
\hline
channel  & LHC 35/pb & bino & wino & Z-rich higgsino & other higgsino  \\
\hline
$\gamma \gamma + \met$ &  \cite{Chatrchyan:2011wc} & \checkmark & & &  \\
$\ell \gamma + \met $ & & & \checkmark & &     \\
jets + $\met$  &  \cite{Khachatryan:2011tk,daCosta:2011qk} & & \checkmark& \checkmark & \checkmark    \\
 $Z(\ell^+ \ell^-) +{\rm jets} + \met$ & & & & \checkmark &    \\
$Z(\ell^+\ell^-)Z(\ell'^+\ell'^-) + \met$  & & & & \checkmark &    \\
\hline
$Z(\ell^+ \ell^-)h(b \bar b) + \met$ & & & & & \checkmark     \\
$h(b \bar b) h(b \bar b) + \met$ & & & & &  \checkmark    \\
$\gamma + h(b \bar b) + \met$ & & & & &  \checkmark    \\
$\gamma+{\rm jets}+\met$ & & \checkmark & \checkmark & & \checkmark   \\
$\ell+{\rm jets}+\met$ & \cite{daCosta:2011hh} & & \checkmark & &    \\
\hline
\end{tabular} \end{center}
\caption{The most promising discovery modes at the LHC for bino, wino and higgsino NLSPs.  In this paper, we have conducted detailed simulations to determine the LHC reach in the first five channels listed  here. }
\label{tab:LHCchannel}
\end{table}

Our minimal spectra will be useful to both theorists and experimentalists for designing and optimizing searches, and reporting and interpreting their results. Below we will discuss the limits from Tevatron searches on our benchmark parameter spaces. Using simple cuts and crude detector simulations, we will also estimate the reach at the LHC in clean final states. 
For easy reference, we summarize here the various Tevatron and LHC search channels that are relevant for GMSB with a promptly-decaying neutralino NLSP\@.  In table~\ref{tab:TeVchannel}, we list the existing Tevatron searches that we find to be most constraining for bino, wino, and $Z$-rich higgsino NLSPs.  In table~\ref{tab:LHCchannel}, we show the LHC search channels that have discovery potential for these scenarios.  We have also included in table~\ref{tab:LHCchannel} the current LHC analyses in these final states. We note that there are still many final states left to be explored. Table~\ref{tab:LHCchannel} should be seen as our personal wishlist for the next round of LHC analyses.

The outline of this paper is as follows.  In section~\ref{sec:bench}, we introduce simple benchmark spectra for studying the colored production of bino, wino, and higgsino NLSPs.  Each spectrum consists of a gluino and the NLSP\@.  We also discuss the production cross-sections and decay branching ratios that will determine the signal rates in the rest of the paper.  Sections~\ref{sec:bino}, \ref{sec:wino}, and~\ref{sec:ZrichHiggsino} contain our main results, where we show the Tevatron limits and LHC reach for our bino, wino, and $Z$-rich higgsino benchmark spectra.  Finally, in section~\ref{sec:OtherHiggsino}, we consider more general higgsino scenarios, with decays to $h$, $\gamma$, and $Z$.  In appendix~\ref{app:nonminspec}, we discuss the consequences of extending our framework to consider a less minimal spectrum, where both a gluino and squarks contribute to the colored production of wino co-NLSPs.

\section{Minimal Spectra for General Neutralino NLSPs}
\label{sec:bench}

In this section, we describe our minimal benchmark parameter spaces for general neutralino NLSPs.  As discussed in the introduction, we will be taking simplifying limits where the NLSP is a gauge eigenstate: either bino, wino or higgsino NLSP\@.   We now highlight several important features of each type of neutralino NLSP, namely the NLSP decay modes and production channels.  For a more detailed discussion, we refer the reader to~\cite{Meade:2009qv}.  

\begin{figure}[b!]
\begin{center}
\includegraphics[scale=0.45]{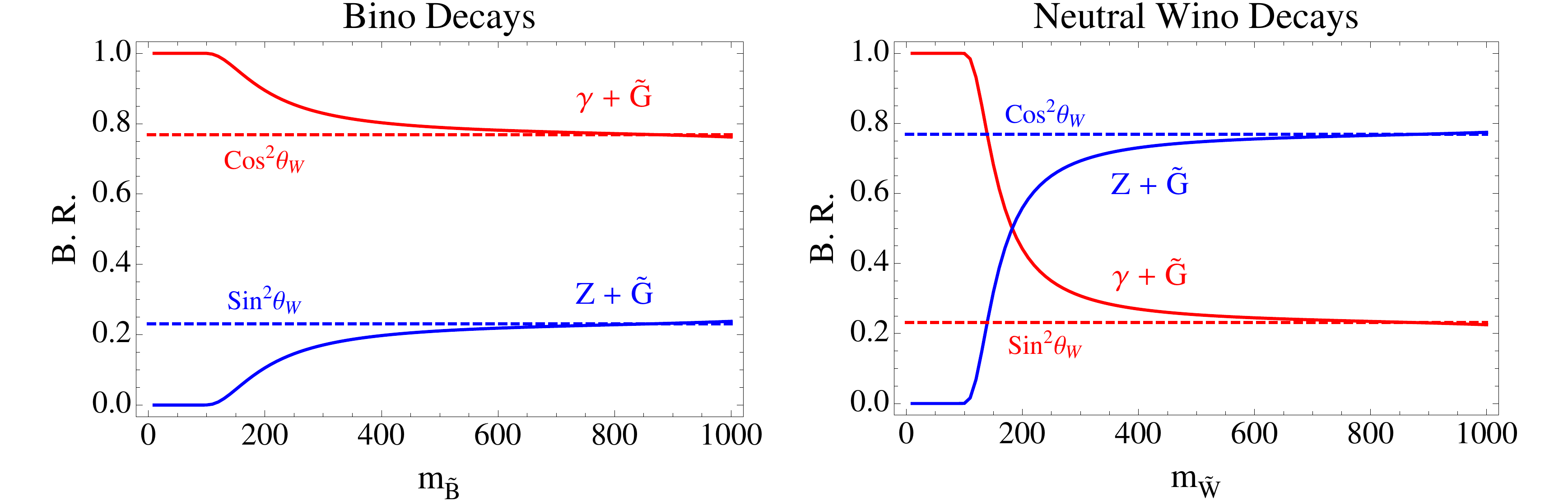}
\end{center}
\caption{The bino and neutral wino NLSP branching fractions to $Z$ or $\gamma$ plus gravitino~\cite{Meade:2009qv}.  The branching fraction is determined by the weak mixing angle, and, at low mass, by the phase space suppression of decays to $Z$'s.}
\label{fig:br}
\end{figure}

A neutralino NLSP decays to $X + \tilde G$, where $X = \gamma, Z, h$, and the different gauge eigenstates are characterized by having different branching fractions to the different  $X$.  The branching fractions of the bino-like and wino-like neutralino NLSP are shown in figure~\ref{fig:br}.  We see that binos dominantly decay to photons with branching fraction $\sim \cos^2 \theta_W$, with a subdominant component to $Z$'s, with branching fraction $\sim \sin^2 \theta_W$.  On the other hand, these branching ratios are flipped for a neutral wino NLSP, which decays mostly to $Z$.  A higgsino NLSP dominantly decays to $Z$ or $h$, with branching ratio that depends on the value of $\tan \beta$ and the sign of $\mu$. Following~\cite{Meade:2009qv}, we will specialize to three cases: (1) the higgsino decays are $Z$ rich at low $\tan \beta$ and positive $\mu$, (2) $h$-rich at low $\tan \beta$ and negative $\mu$, and (3) roughly evenly mixed between $Z$ and $h$ at moderate to large $\tan \beta$.  (The above discussion applies to the entire GGM parameter space, but it is important to keep in mind that more complicated frameworks, such as the presence of multiple supersymmetry breaking sectors, can lead to different NLSP branching fractions~\cite{Thaler:2011me}.)

We note that when the NLSP is mostly wino, there is generically a very small splitting between the charged and neutral wino~\cite{Martin:1993ft}.  When this happens, the three-body decay to the neutral wino becomes squeezed out, and the charged wino prefers to decay directly to $W^\pm$ and a gravitino~\cite{Meade:2009qv}. In other words, the neutral and charged winos become co-NLSPs, and the final states contain $W$'s, along with $Z$'s and $\gamma$'s.  On the other hand, the splitting between the charged and neutral higgsinos is generically larger, such that only the lightest neutralino decays directly to the gravitino.

\begin{figure}[t!]
\begin{center}
\includegraphics[scale=0.65]{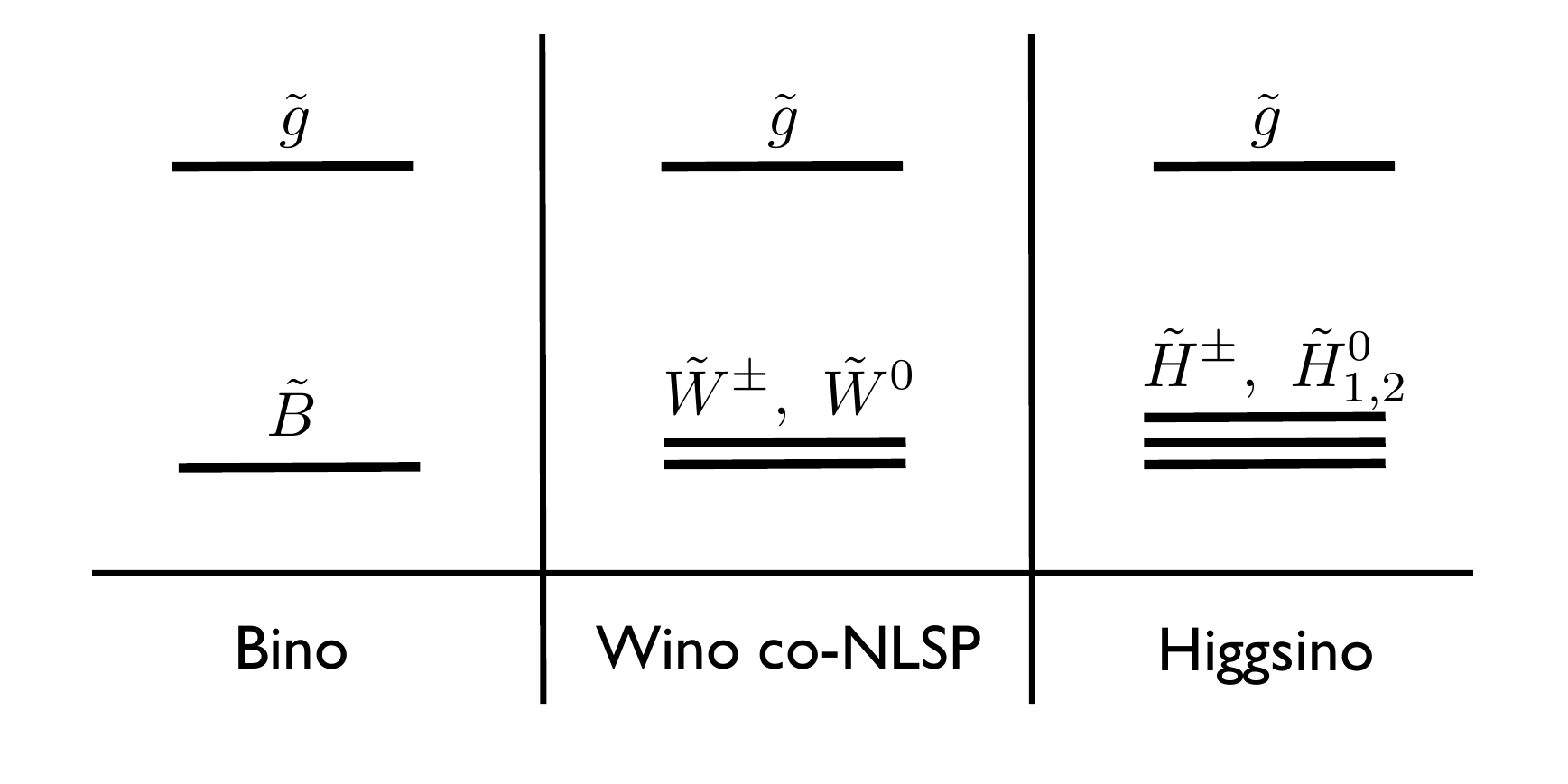}
\end{center}
\caption{The three minimal parameter spaces for general neutralino NLSPs that we study in this paper.  We consider a gluino that is heavier than a bino, winos, or higgsinos, with the other superpartners decoupled.}
\label{fig:schematic}
\end{figure}

Our simplified spectra are shown in figure~\ref{fig:schematic}.  Basically, we consider varying the gluino mass $M_3$ and the NLSP mass ($M_1$, $M_2$ or $\mu$ depending on whether the NLSP is bino, wino or higgsino, respectively). All other states are decoupled (their masses are set to 1.5 TeV), since they do not play an important role in the signatures of interest.  The MSSM higgs sector is taken to be in the decoupling limit, with the SM higgs mass set to $m_h=120$ GeV\@. We always fix the NLSP decay length to be prompt, $c\tau_{NLSP}=0.1$~mm.  For the higgsino simplified model, we specialize to the $Z$ rich case, fixing $\tan \beta = 2$ and $\mu>0$.  In section~\ref{sec:OtherHiggsino}, we will also consider scenarios with $h$-rich and mixed higgsinos. 

Our benchmarks yield the minimal spectrum with strong production cross-sections at the LHC and inclusive signatures from NLSP decays to the gravitino. As long as searches are designed to look inclusively for the NLSP decay, including additional states between the gluino and the NLSP should not make a difference. We have verified that including squarks in addition (or in place) of gluinos does not affect the conclusions very much -- the predominant effect being a modified colored production cross-section.  We will consider an example with both a gluino and squarks in appendix~\ref{app:nonminspec}.  We emphasize that these minimal parameter spaces do correspond to physical models, since the entire GGM parameter space was covered by a perturbative messenger model in~\cite{Buican:2008ws}.

\begin{figure}[t!]
\begin{center}
\includegraphics[scale=0.4]{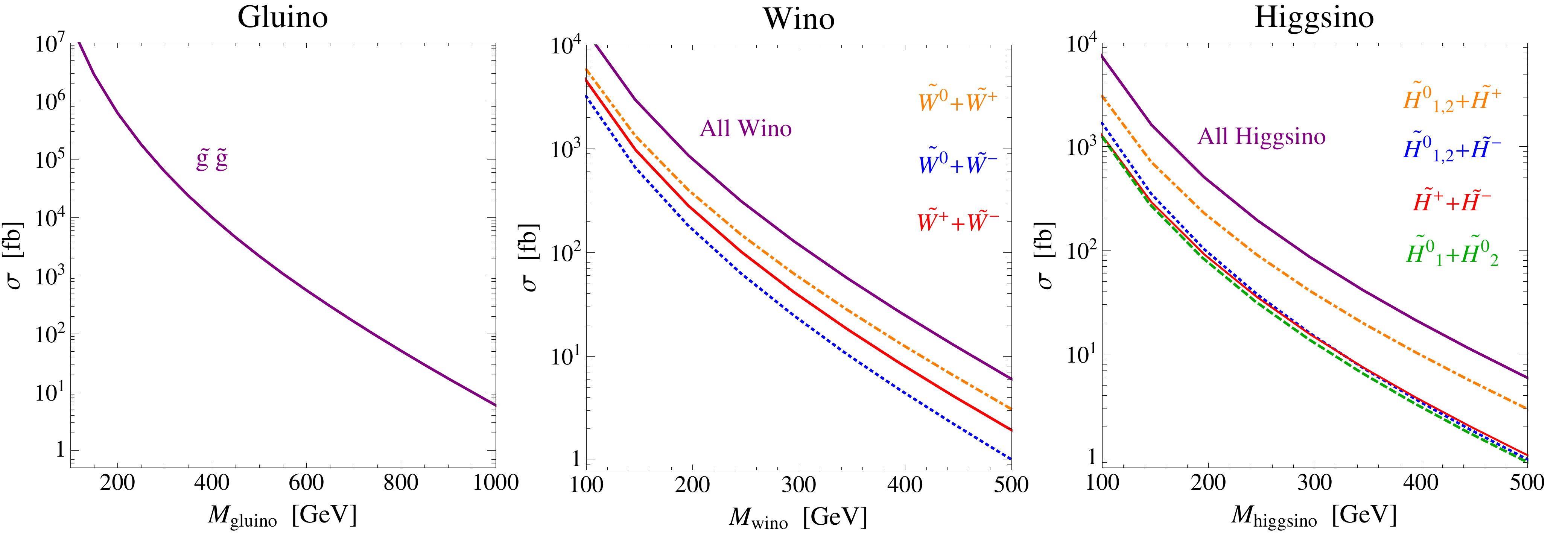}
\end{center}
\caption{The NLO production cross-sections at 7 TeV LHC relevant for our benchmark models~\cite{Beenakker:1996ch,Beenakker:1999xh}.  For each model, we consider the colored production of gluinos, shown to the {\it left}.  For the wino and higgsino benchmarks, there are also electroweak production modes, shown to the {\it center}, and {\it right}, respectively.}
\label{fig:cross}
\end{figure}

Our simplified spectra are characterized by several types of SUSY production, with NLO cross-sections shown in figure~\ref{fig:cross}.  For each benchmark, there is colored gluino production, with rate set by the gluino mass.  For the wino and higgsino NLSPs, there is also the possibility of direct NLSP production with electroweak cross-sections. 
While the Tevatron currently has the advantage here because of its much larger dataset, we will see below that the LHC will have sensitivity to electroweak production with $\gtrsim 1-5$~fb$^{-1}$.

\section{Bino NLSP}
\label{sec:bino}

\subsection{Tevatron Limits}
\label{sec:BinoTeV}

We begin by specializing to pure bino NLSP, and determining the Tevatron limit on colored SUSY production. Note that for both Tevatron and LHC, the direct bino production cross-section is negligible, so all the limits on this scenario will come from production of heavier states, such as gluinos in our case.

\begin{figure}[t]
\begin{center}
\includegraphics[scale=0.55]{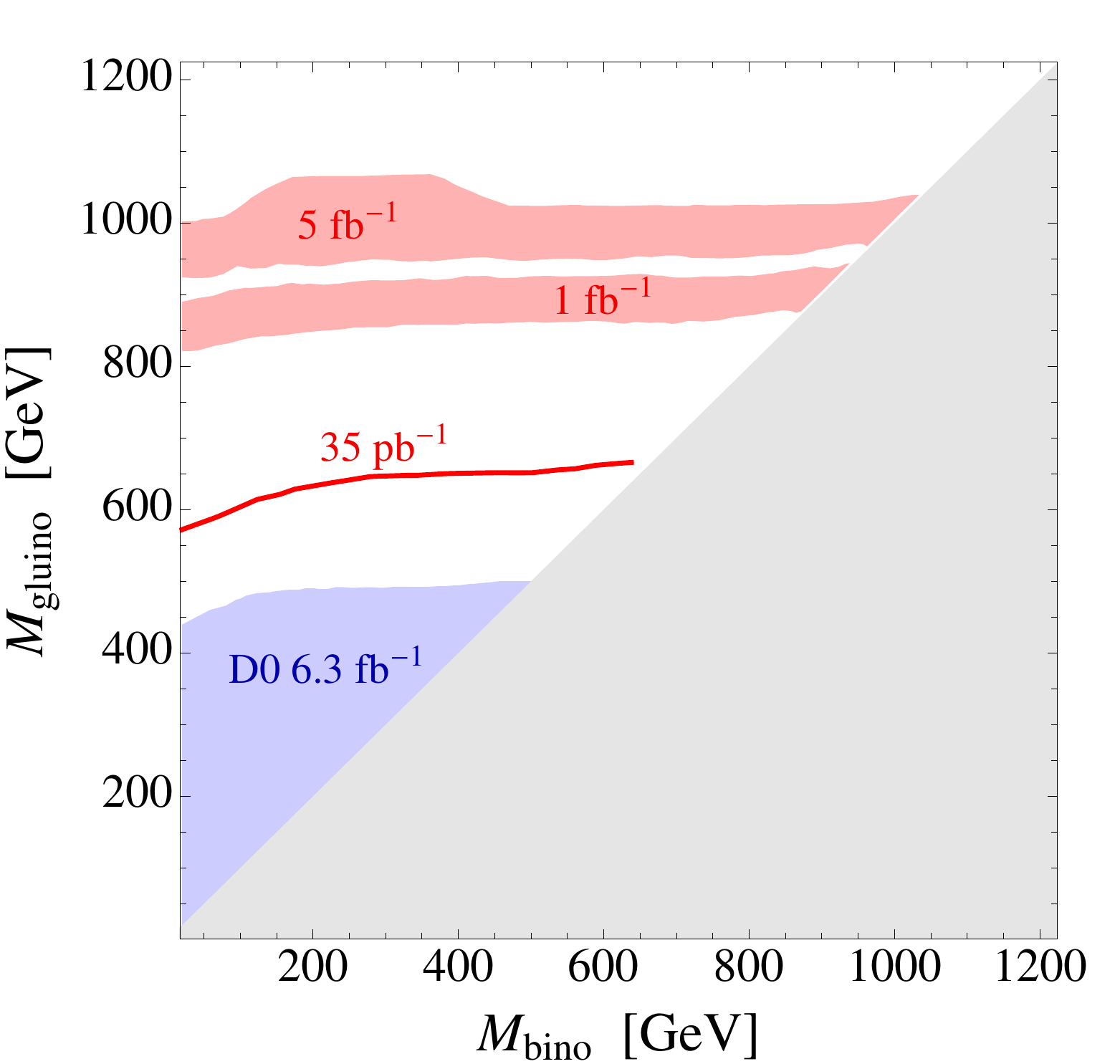}
\end{center}
\caption{The Tevatron limit (blue) and 7 TeV LHC reach (red) for our bino NLSP benchmark.  We allow the background to range from 1-10~fb.  Backgrounds at this level do not affect the 35~pb$^{-1}$ reach, but lead to a reach that varies within the shaded bands for 1 and 5~fb$^{-1}$.  The gray triangular region corresponds to gluino NLSP, $m_{\tilde g} < m_{\tilde B}$, which we do not consider in this paper.}
\label{fig:binolimits}
\end{figure}

As discussed above, the dominant NLSP decay mode is $\tilde B \rightarrow \gamma + \tilde G$. The strongest limit comes from the  $\gamma \gamma + \met$ final state, as described in table~\ref{tab:TeVchannel}.  This is the leading channel because of the large branching ratio of about $(\cos^2 \theta_W)^2 \sim 0.6$, and low SM background, which is dominated by QCD with fake missing energy, and a combination of real and fake photons.  The most recent search in this channel was carried out by D\O~with 6.3~fb$^{-1}$~\cite{D0ggmet}.  

We determined the limit from the D\O~$\gamma \gamma + \met$ search on the bino NLSP parameter space, defined in section~\ref{sec:bench}, by simulating the efficiency of the search on our signal.  Here, and throughout this paper, we use Pythia 6.4 for event generation~\cite{Sjostrand:2006za}, and PGS~4 for detector simulation~\cite{PGS}.  For the Tevatron, we use LO pythia cross-sections throughout. (This is just for simplicity -- including K-factors will make very little difference on the limit contours in the mass plane, since the K-factors at the Tevatron are generally at most $\sim 1.5$ even for colored production, and since the cross-sections are rapidly falling functions of mass.) The limit is depicted by the blue shaded region in figure~\ref{fig:binolimits}.  We see that the Tevatron excludes $m_{\tilde g} \lesssim 500$~GeV\@. The limit mostly depends on the gluino mass, which sets the production cross-section, but it does weaken slightly at lighter bino masses, $\tilde m_{\tilde B} \lesssim 100$~GeV\@.  With such light binos, there is less energy available for the photons and gravitinos, leading to a lower acceptance.

Since there are always at least two jets in every event coming from gluino decay, we have also checked the Tevatron limit from jets plus missing energy, using the D\O~search with 2.1~fb$^{-1}$~\cite{D0jetsmet}.  This search always sets a much weaker limit than $\gamma \gamma + \met$ on bino NLSP, due to larger backgrounds.

\subsection{LHC Reach}
\label{sec:BinoLHC}

 \begin{table}[t!]
\begin{center}
\begin{tabular}{|c|}
\hline
  LHC  $\gamma\gamma+\met$ \\ 
     \hline 
   $ N_{\gamma}  \geq 2$ \\
 $p_{T,\gamma} > 50$ GeV \\ 
 $ | \eta_{\gamma} | < 1.5$  \\
 $\met > 100$ GeV   \\
 \hline
\end{tabular}
\caption{Event selection criteria for a hypothetical $\gamma \gamma + X+ \met$  search
at the LHC. 
}
\label{ggcutstable}
\end{center}
\end{table}

At the LHC, $\gamma \gamma+\met$ remains the best discovery mode for bino NLSPs.  In order to estimate the discovery reach, we have chosen simple cuts, shown in table~\ref{ggcutstable}.  These cuts are designed to look inclusively for two hard, central, isolated\footnote{For the LHC reach estimates in this paper, we have modified the photon and lepton isolation definitions of PGS to demand that the extra calorimeter $E_T$, in a $\Delta R < 0.4$ cone, is less than 10\% of the $E_T$ ($p_T$) of the photon (lepton).} photons, together with a missing energy requirement of $\met > 100$~GeV\@.  It is difficult to estimate the background of these cuts, since it is dominated by QCD with fake missing energy. 
 Instead of attempting to simulate the background, we estimate the exclusion reach of the search using the background range of 1 - 10~fb.  We believe this estimate to be reasonable, given the existing MC studies and actual searches by the LHC collaborations (see e.g.~\cite{CMSdigamma,Chatrchyan:2011wc}).

The expected exclusion contours are shown in figure~\ref{fig:binolimits}, with luminosities of 35~pb$^{-1}$, 1~fb$^{-1}$, and 5~fb$^{-1}$.  Here, and throughout the rest of the paper, we use Prospino 2.1 for NLO SUSY production cross-sections at the LHC~\cite{Beenakker:1996ch, Beenakker:1999xh}.  The expected exclusion contours correspond to 95\% confidence level limits using the CL$_s$ statistic~\cite{Junk:1999kv}, assuming the null hypothesis that the number of data counts will be equal to the expected background.  The reach with 35~pb$^{-1}$ corresponds to a gluino mass of about 600 GeV\@.  We note that CMS has already published a search in this channel, and the limit they set is close to our estimate~\cite{Chatrchyan:2011wc}.  Similarly to the Tevatron limit, the reach weakens slightly at lower bino mass, $m_{\tilde B} \lesssim 100$~GeV, due to less energy available for the photons and gravitinos.  Looking ahead to the 1-5~fb$^{-1}$ that are expected from the rest of the 7 TeV run, the reach is spectacular, probing gluinos up to about 1 TeV\@.

\section{Wino co-NLSP}
\label{sec:wino}

\subsection{Tevatron Limits}
\label{sec:WinoTeV}

\begin{figure}[t!]
\begin{center}
\includegraphics[scale=0.5]{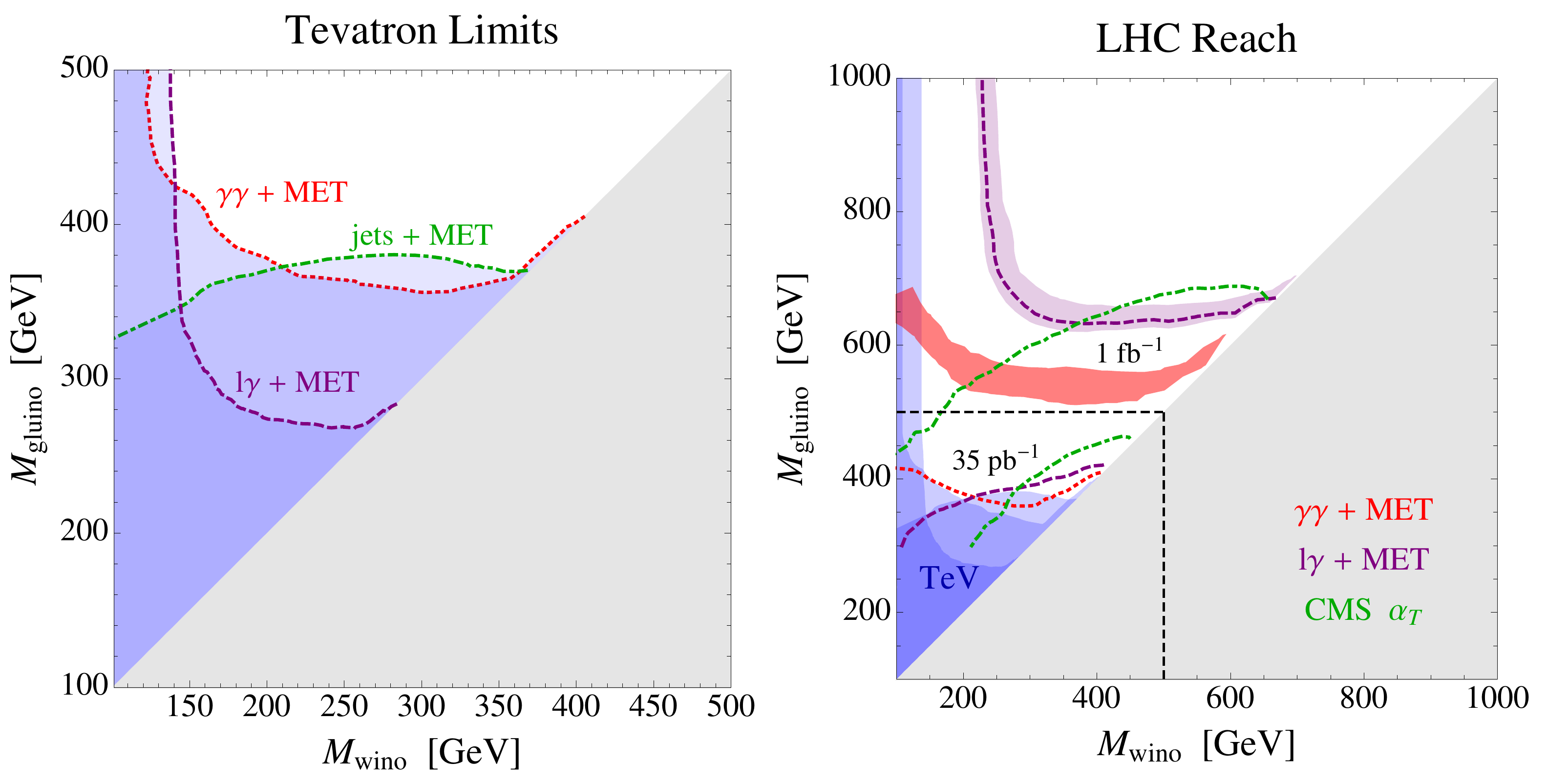}
\end{center}
\caption{The Tevatron limit ({\it left}) and LHC reach ({\it right}), for our wino co-NLSP benchmark.  At the Tevatron, we see that the $l^\pm \gamma$ (purple), $\gamma \gamma$ (red), and jets plus MET (green) searches are complementary, with each search dominating the limit in different parts of the gluino/wino mass plane.  At the LHC, we show the reach of our proposed $\gamma \gamma$ and $l^\pm \gamma$ searches, compared to the CMS $\alpha_T$ search, with 35~pb$^{-1}$ and 1~fb$^{-1}$.  For the $\gamma \gamma$ search, we allow the background to vary from 1 - 10~fb.  For the $l^\pm \gamma$ search, we vary the background from 1/2 to 2 times our estimate of 1.43~fb.  Backgrounds at this level do not matter at 35~fb$^{-1}$, but they lead to the purple band at 1~fb, where the dotted line corresponds to a background of 1.43~fb.}
\label{fig:winoreach}
\end{figure}

Now we consider the reach of wino co-NLSPs at the Tevatron.  We recall from the discussion in section~\ref{sec:bench} that both the charged and neutral winos are co-NLSPs, so both prefer to decay directly to gravitinos.  Several different production modes and final state channels are relevant for wino co-NLSPs.  

\begin{itemize}

\item {\it Direct wino production.} Winos can be produced directly in the combinations $\tilde W^+ \tilde W^-$ (through an s-channel $Z$) and $\tilde W^0 \tilde W^\pm$ (through an $s$-channel $W^\pm$).  The former leads to $W^+ W^-$ in the final state, and the latter leads to a $W^\pm$ and a $Z/\gamma$.  

\item {\it Colored production.} When gluinos are produced, each gluino can decay to either a charged or neutral wino, leading to jets plus $\tilde W^0 \tilde W^\pm$, $\tilde W^+ \tilde W^-$, $\tilde W^0 \tilde W^0$, or $\tilde W^\pm \tilde W^\pm$. Note that the final two possibilities are unique to colored production, and they can lead to $\gamma \gamma$ and same-sign lepton final states. In our benchmark, the gluino has an $\mathcal{O}(1)$ branching fraction to decay to both the charged and neutral wino.  For example, fixing the gluino (wino) mass to 700 (300) GeV, there is a 35\% branching ratio to the neutral wino, and a 65\% branching ratio to the charged wino.

\end{itemize}

Due to the many possibilities discussed above, we have found that several different channels searched for at the Tevatron place complementary limits on wino co-NLSPs.  The most constraining channels are listed in table~\ref{tab:TeVchannel}: missing energy with $l^\pm \gamma$, $\gamma \gamma$, or jets.  We show our estimate of the limits from these searches on our wino co-NLSP parameter space, on the left side of figure~\ref{fig:winoreach}.  For the $l^\pm \gamma$ search, we have raised the missing energy cut from 25 to 50 GeV using the $\met$ distributions in~\cite{cdflgX}, as this has been shown to improve the signal significance~\cite{Meade:2009qv}.  We see that the three searches are complementary, with jets$+\met$ dominating at high wino mass; $l+\gamma+\met$  winning at low wino mass, because it probes direct electroweak production in addition to strong production; and $\gamma \gamma+\met$  setting the strongest limit at intermediate wino masses.  We note that we have also checked the Tevatron limit from same-sign dileptons, using the 1~fb$^{-1}$ search by CDF~\cite{CDFSSdilep}, and the limit is weaker than the channels discussed above, due to the relatively low branching ratio.

\subsection{LHC Reach}
\label{sec:WinoLHC}

\begin{figure}[t!]
\begin{center}
\includegraphics[scale=.5]{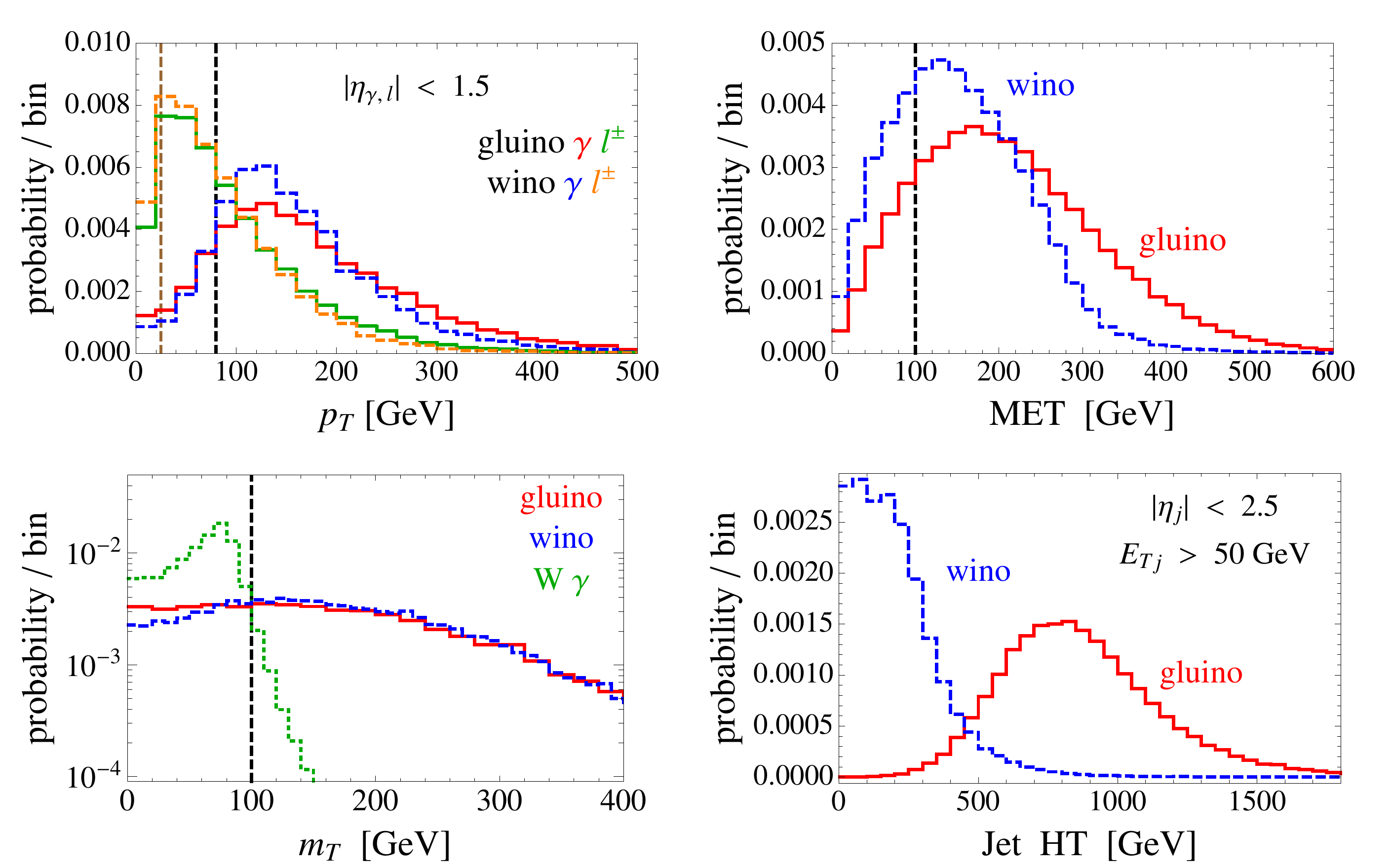}
\end{center}
\caption{A comparison of electroweak/wino versus colored/gluino production for our benchmark. We have taken $m_{\tilde W}=300$~GeV and $m_{\tilde g} =700$~GeV, points where the overall colored and electroweak cross-sections are comparable, at 160 and 120~fb, respectively.  The {\it upper left} plot shows the lepton and photon $p_T$ spectra, the {\it upper right} plot shows the missing energy, the {\it lower left} plot shows the $\ell-\met$ transverse mass (with SM $W+\gamma$ also included for comparison), and the {\it lower right} plot shows the $H_T = \Sigma |E_T|$ of jets.  The cuts of our proposed search are showed as vertical dashed lines.}
\label{fig:winokindist}
\end{figure}

We consider the LHC reach to discover wino co-NLSPs, using the channels listed in table~\ref{tab:LHCchannel}: $l^\pm \gamma$, $\gamma \gamma$, or jets plus missing energy.  These are the same channels that we found to set the strongest limits at the Tevatron.  For $\gamma \gamma + \met$, we use the same cuts as for bino NLSP, described in table~\ref{tab:LHCchannel}.  For jets$+\met$, we take as an example the CMS $\alpha_T$ search conducted with 35~pb$^{-1}$~\cite{Khachatryan:2011tk}, whose performance we extrapolate to higher luminosities.

In order to estimate the LHC reach in the $l^\pm \gamma + \met$ final state, we use the cuts in table~\ref{lgcutstable}.  We require at least one hard, central, isolated lepton and photon,  large missing energy, and large transverse mass between the hardest lepton and the missing energy. These cuts were crudely optimized to distinguish between background and signal. Some kinematic distributions for colored vs.\ electroweak production are shown in figure~\ref{fig:winokindist}.  We see that the transverse mass cut is extremely effective at removing the $W^\pm \gamma$ background, while keeping most of the signal. In the other kinematic distributions, we see that gluino production leads to a slightly harder MET spectrum, similar lepton and photon kinematics, and significantly more hadronic energy manifested in the jet $H_T$. In order to be sensitive to direct wino production, we found that a hard cut on $H_T$ was counterproductive.

\begin{table}[t!]
\begin{center}
\begin{tabular}{|c|c|}
\hline 
 \multicolumn{2}{|c|}{LHC  $\ell\gamma+\met$} \\ 
   \hline 
   \hspace{1cm} $ N_{\gamma},\,N_{\ell}  \ge 1$  \hspace{1cm}  &  \hspace{1cm} $ | \eta_{\gamma} |,\, |\eta_{\ell}| < 1.5$  \hspace{1cm}  \\
 $p_{T,\gamma} > 80$ GeV & $\met > 100$ GeV   \\ 
 $p_{T,\ell} > 25$ GeV &$m_{T}(\ell,\met)>100$ GeV \\ 
 \hline 
\end{tabular}
\caption{Event selection criteria for a hypothetical $\ell \gamma + X+ \met$  search
at the LHC. 
}
\label{lgcutstable}
\end{center}
\end{table}

\begin{table}[t]
\begin{center}
\begin{tabular}{|c|c|c|}
\hline
Background & Cross section (fb) & Cross section after cuts (fb) \\
\hline
$W(\ell\nu)+\gamma+{\rm jets}$ & 711 & 0.35 \\
\hline
$tt+\gamma$ & 63.4 & 0.59 \\
\hline
$tt(\ell\ell)$ & 9460 & 0.49\\
\hline
\hline
{\rm Total} & -- & 1.43\\
\hline
\end{tabular}
\caption{SM background rates before and after cuts for a hypothetical $\ell+\gamma+\met$ search at the LHC\@. These were generated using MadGraph with a $p_T>80$ GeV cut on the photon. They are LO only. The $W+\gamma$~+~jets is a fully-matched $0+1+2$ jet sample with a jet matching scale of 15 GeV.
\label{tab:SMwgbg}}
\end{center}
\end{table}

We have estimated the SM background rate for these cuts, which is dominated by $W \gamma$ + jets, $t \bar t \gamma$, and $t \bar t$.  For our background estimate, which is shown in table~\ref{tab:SMwgbg}, we use Madgraph~4.4~\cite{Alwall:2007st} for event generation and PGS~4~\cite{PGS} for detector simulation.  For $W \gamma$, we use a matched sample with up to 2 jets, with a matching scale of 15 GeV\@.  The $t \bar t$  background corresponds to doubly leptonic tops, where an electron fakes a photon.  This fake rate is dominated by the probability of losing the track pointing to the calorimeter, which we estimate to be 2\%\footnote{We thank Y. Gershtein for help with this point.}.  Our total background estimate is 1.43~fb, but to allow for systematic errors in this estimate, we consider a band of background estimates ranging from $(0.5-2)\times 1.43$~fb.

The estimated reach for these three channels is shown to the right of figure~\ref{fig:winoreach}, with integrated luminosities of 35~pb$^{-1}$ and 1~fb$^{-1}$.  We see that jets plus missing energy provides the best reach at high wino mass, while the $\gamma \gamma$ and $l^\pm \gamma$ have stronger reach at intermediate wino mass.  We also see that by 1~fb$^{-1}$, the LHC will surpass the Tevatron reach for electroweak production, and the $l^\pm \gamma$ channel will probe the direct production of wino co-NLSPs up to masses of 250 GeV.  

\begin{figure}[t!]
\begin{center}
\includegraphics[scale=.6]{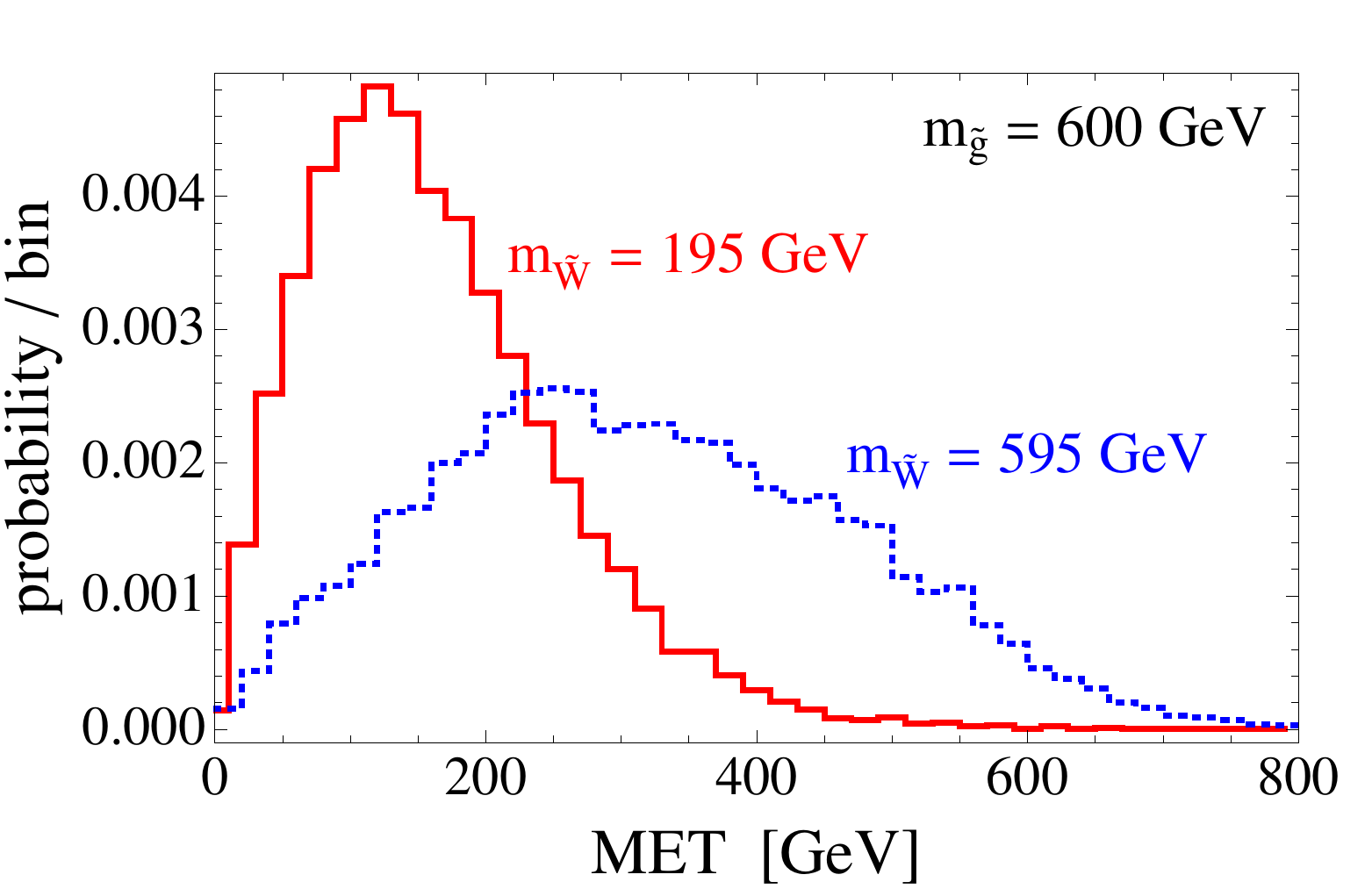}
\end{center}
\caption{The  $\met$ distributions of our wino co-NLSP benchmark, fixing $m_{\tilde g} = 600$~GeV, and comparing heavy (595 GeV) versus light (195 GeV) winos.  We see that the heavier wino leads to a much harder $\met$ distribution, which explains why jets$+\met$ is more powerful at heavier wino mass.  }
\label{fig:winomet}
\end{figure}

A few more comments about the sensitivity of jets+$\met$ to wino NLSPs are in order. First, the dramatic weakening of the jets+$\met$ search at lower wino masses is interesting. The reason for this behavior is that at lower wino mass, there is significantly less energy available for the gravitinos, and therefore less missing energy.  This is shown in figure~\ref{fig:winomet}, where the $\met$ distributions for two different wino masses and the same gluino mass are overlaid.
Second, we note that at high wino mass, where the $\met$ is highest, the jet energy comes from the electroweak $Z$ and $W$ decays, instead of the gluino decay.  Finally, we comment that the success of jets+$\met$ is model dependent, because such searches usually veto leptons in order to fight the SM background from $W$ plus jets.  Intermediate sleptons in the spectra can give leptons in the final state that activate these vetoes.  In this situation, more inclusive $l^\pm \gamma$ and $\gamma \gamma$ searches are more effective.

Finally, let us also mention two more signatures that would be interesting to explore, and which we expect to have competitive sensitivity to wino co-NLSPs,  compared to the final states considered here. These are $\gamma+{\rm jets}+\met$ and $\ell+{\rm jets}+\met$. The point is that the $\gamma\gamma+\met$ and $\ell\gamma+\met$ final states considered here, while very clean, are branching ratio suppressed. For instance, to go to $\gamma\gamma+\met$, one pays a branching fraction price of ${\rm Br}(\tilde g\to \tilde W^0+X)^2\times {\rm Br}(\tilde W^0\to \gamma+\tilde G)^2 \sim 1\%$. Thus requiring only a single photon or a single lepton will enhance the signal rate considerably. By cutting harder on jets and $\met$, perhaps one can suppress SM backgrounds while keeping the signal rate high. ATLAS has recently put out a 35~pb$^{-1}$ search for $\ell+{\rm jets}+\met$, and it would be interesting to derive its constraints and projected reach in our wino co-NLSP parameter space.  There is so far no published 35~pb$^{-1}$ LHC search for $\gamma+{\rm jets}+\met$, but we suggest that such a search be carried out.

\section{Z-rich Higgsino NLSP}
\label{sec:ZrichHiggsino}

\subsection{Tevatron Limits}
\label{sec:HiggsinoTeV}

\begin{figure}[t!]
\begin{center}
\includegraphics[scale=0.5]{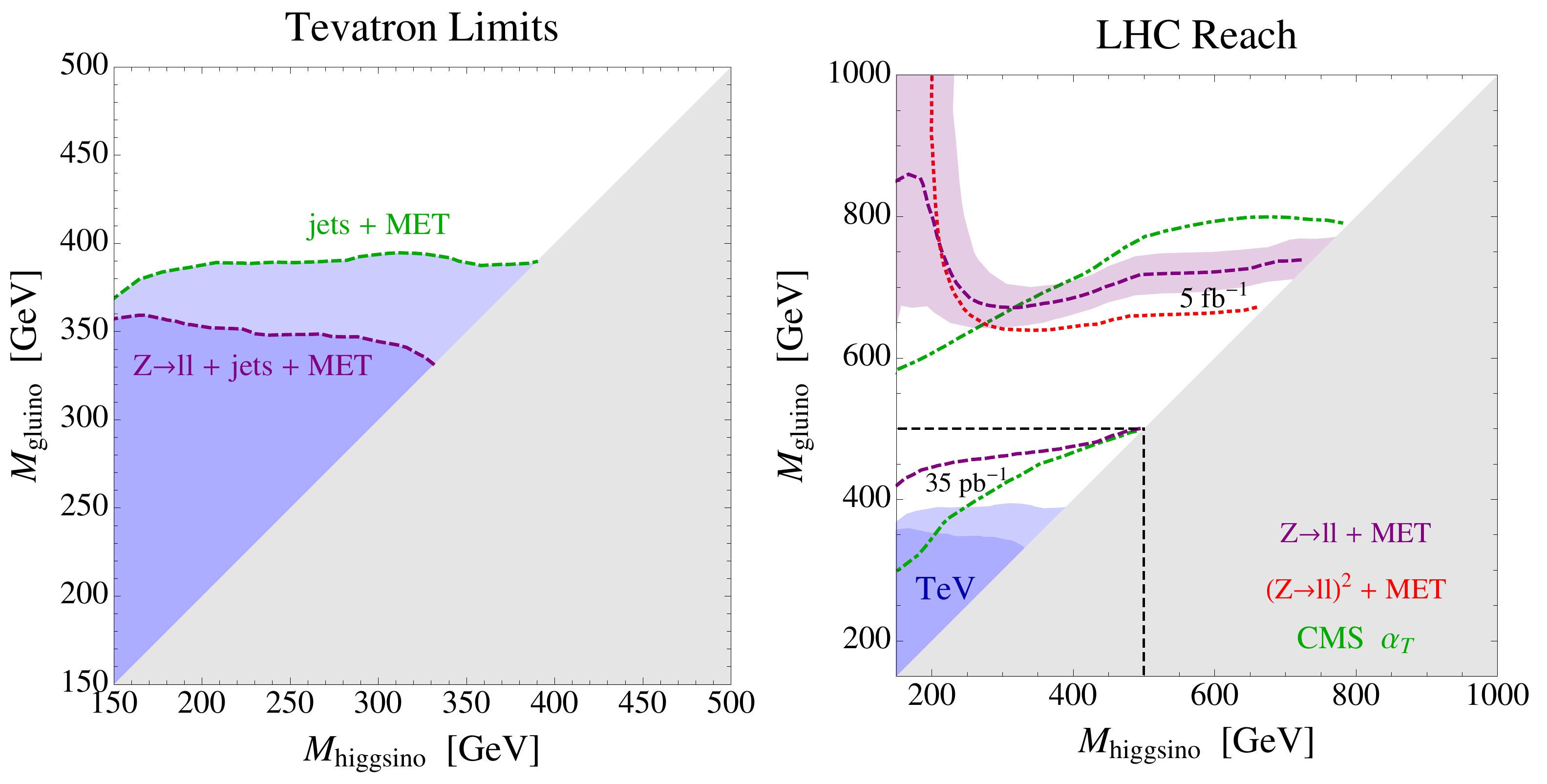}
\end{center}
\caption{The Tevatron limit ({\it left}) and LHC reach ({\it right}), for our higgsino NLSP benchmark.  At the Tevatron, we find that jets plus missing energy (green) sets a stronger limit than the search for a leptonic $Z$ plus jets and missing energy.  At the LHC we compare the reach with our proposed single and doubly leptonic $Z$ searches, to the CMS $\alpha_T$ search.  At 35~pb$^{-1}$, the doubly leptonic Z search does not improve upon the Tevatron limits, so is not shown.  We then show the limit with 5~fb$^{-1}$, which roughly corresponds to the luminosity required to see the electroweak branch at low $m_{\tilde H}$.  The purple band corresponds to 1/2 and 2 multiplied by our background estimate of 28~fb$^{-1}$, shown as a dashed line.}
\label{fig:higgsinoreach}
\end{figure}

In this section, we address the Tevatron limits on a higgsino NLSP that dominantly decays to $Z$ bosons.  As discussed in section~\ref{sec:bench}, the lightest neutral higgsino decays dominantly to $Z$'s when $\tan \beta$ is low and $\mu$ is positive.  The strongest Tevatron limits are listed in table~\ref{tab:TeVchannel}: jets$+\met$ (the jets here come from both the gluino decay and $Z$ decays), and a CDF search for $Z(e^+ e^-)+{\rm jets}+\met$, where the invariant mass of the two jets are required to fall within 60 to 95 GeV\@. Our estimate of the limits of these searches is shown to the left of figure~\ref{fig:higgsinoreach}.  We see that the D\O~jets$+\met$ search beats the search for $Z\rightarrow e^+ e^-$  by a small margin.  We also see that both limits are mostly independent of higgsino mass, and correspond to a limit on gluino mass of about 350-400 GeV\@.  The electroweak production of higgsinos has not yet been constrained by the Tevatron~\cite{Meade:2009qv}.

We note that higgsinos introduce an important difference in the topology of the gluino decays.  In the previous bino and wino benchmarks, the gluino decayed three-body to the neutralino and two jets, through an off-shell squark.  But the higgsino couples predominantly to heavy flavor, where the mass of the top can squeeze out the three-body decays.  In this regime, the dominant gluino decay is a one-loop two body decay, $\tilde g \rightarrow g \, \tilde H_{1,2}$.  We use SDECAY~\cite{Muhlleitner:2003vg} to compute the relative branching ratios of the two and three-body decay modes.

\subsection{LHC Reach}
\label{sec:HiggsinoLHC}

We now consider the LHC reach for a higgsino NLSP that decays predominantly to $Z$'s.  We suggest two dedicated searches for this final state. One requires one leptonic $Z$ plus jets plus missing energy, and another  looks for $ZZ\to 4\ell$ plus missing energy. The latter is an extremely clean channel, but it suffers from a suppressed signal rate due to branching ratio suppression ${\rm Br}(Z\to \ell^+\ell^-) \sim 0.06^2$. As shown in table~\ref{tab:LHCchannel}, we also compare the reach of these two final states to the performance of jets plus missing energy, where we again use the cuts of the CMS $\alpha_T$ search.

The cuts for our suggested search for one leptonic $Z$ plus jets and missing energy  are shown in table~\ref{ZHTcutstable}. We demand that two opposite-sign same flavor leptons reconstruct the $Z$ mass, we require missing energy above 100 GeV, and we require $H_T$ above 100 GeV, where here the $H_T$ includes both jet and lepton $p_T$ (jets with $p_T>10$ GeV and $|\eta|<2.5$, and leptons with $p_T>10$ GeV and  $|\eta|<1.5$).  The dominant background is dileptonic $t \bar t$ where the two leptons happen to fall within the $Z$ mass window.  We estimate this background rate to be 21~fb, including an NLO K-factor~\cite{Campbell:2010ff}.  The second important background is leptonic decays of dibosons, $ZZ$ and $ZW$, which contribute a total of 7~fb at LO\@.  In order to estimate the reach, we take the background to be the sum of these two sources (so 28~fb in total). In order to account for systematic uncertainty, we consider a range of backgrounds from 0.5 to 2 times this estimate.

For two leptonic $Z$'s plus MET we use the cuts shown in table~\ref{ZZ4lcutstable}.  Here we require at least four isolated leptons with moderate $p_T$ and $\met$ cuts. For this channel, we assume that the background is zero.

\begin{table}[t!]
\begin{center}
\begin{tabular}{|C{6cm}|C{6cm}|}
\hline
 \multicolumn{2}{|C{12cm}|}{LHC  $Z(\ell^+\ell^-)+H_T+\met$ }\\ 
   \hline 
 $ N_{\ell^+},\,N_{\ell^-}  \ge 1$ &   $m_{inv}(\ell^+\ell^-)\in (85,\,95)$ GeV  \\
 $p_{T,\ell^\pm} > 20$ GeV  &  $\met > 100$ GeV   \\
 $ | \eta_{\ell^\pm} | < 1.5$  &   $H_{T}>100$ GeV\\
 \hline
\end{tabular}
\caption{Event selection criteria for a hypothetical $Z(\ell^+\ell^-)+H_T + \met$  search
at the LHC. 
}
\label{ZHTcutstable}
\end{center}
\end{table}

\begin{table}[t]
\begin{center}
\begin{tabular}{|c|}
\hline
 LHC  $Z(\ell^+\ell^-)Z(\ell'^+\ell'^-)+\met$ \\ 
   \hline 
 $N_{\ell}\ge 4$  \\
 $p_{T,\ell} > 10$ GeV  \\
 $ | \eta_{\ell} | < 2.5$  \\
  $\met > 50$ GeV   \\
 \hline
\end{tabular}
\caption{Event selection criteria for a hypothetical $4\ell+\met$  search at the LHC.
}
\label{ZZ4lcutstable}
\end{center}
\end{table}

The estimated exclusion reach for $Z$-rich higgsino NLSP is shown in figure~\ref{fig:higgsinoreach}.  At 35~pb$^{-1}$ we find that the search for one leptonic $Z$ sets the strongest limit.  The search for two leptonic $Z$'s does not yet surpass the Tevatron limit, due to the small branching fraction, so the reach is not shown.  We also show the reach for all three searches with 5~fb$^{-1}$.  We find that jets plus missing energy sets the strongest limit at high higgsino mass.  This is because large higgsino masses lead to more energy in the gravitinos and therefore higher missing energy.  Meanwhile, the searches for one or two leptonic $Z$'s are stronger at low higgsino mass, and we find that the electroweak production branch may be visible with 5~fb$^{-1}$.

A couple of comments about existing LHC searches are in order. There is a 35~pb$^{-1}$ ATLAS search for multileptons plus missing energy~\cite{ATLASmultilepnote}. However it (like many multilepton searches before it!) vetoes on $Z$'s in an effort to reduce the SM background. Such multilepton searches will completely miss the $ZZ\to 4\ell$ plus $\met$ final state discussed here. Second, there is a 35~pb$^{-1}$ CMS search for OS dileptons plus jets plus missing energy~\cite{Chatrchyan:2011bz}. In one search strategy, they again veto on $Z$'s, which renders that part of the search irrelevant for our purposes. In a second search strategy, they do not veto on $Z$'s, but instead require extremely hard $\met$ and $H_T$. This is also not optimized for the Z-rich higgsino scenario. Such hard $\met$ and $H_T$ cuts will have a drastic effect on our signal acceptance, especially in the electroweak production branch. We believe a more optimized approach is to instead  adopt a tight $Z$-mass window as in table~\ref{ZHTcutstable}.

\section{Other Higgsino Types}
\label{sec:OtherHiggsino}

\begin{figure}[b!]
\begin{center}
\includegraphics[scale=0.55]{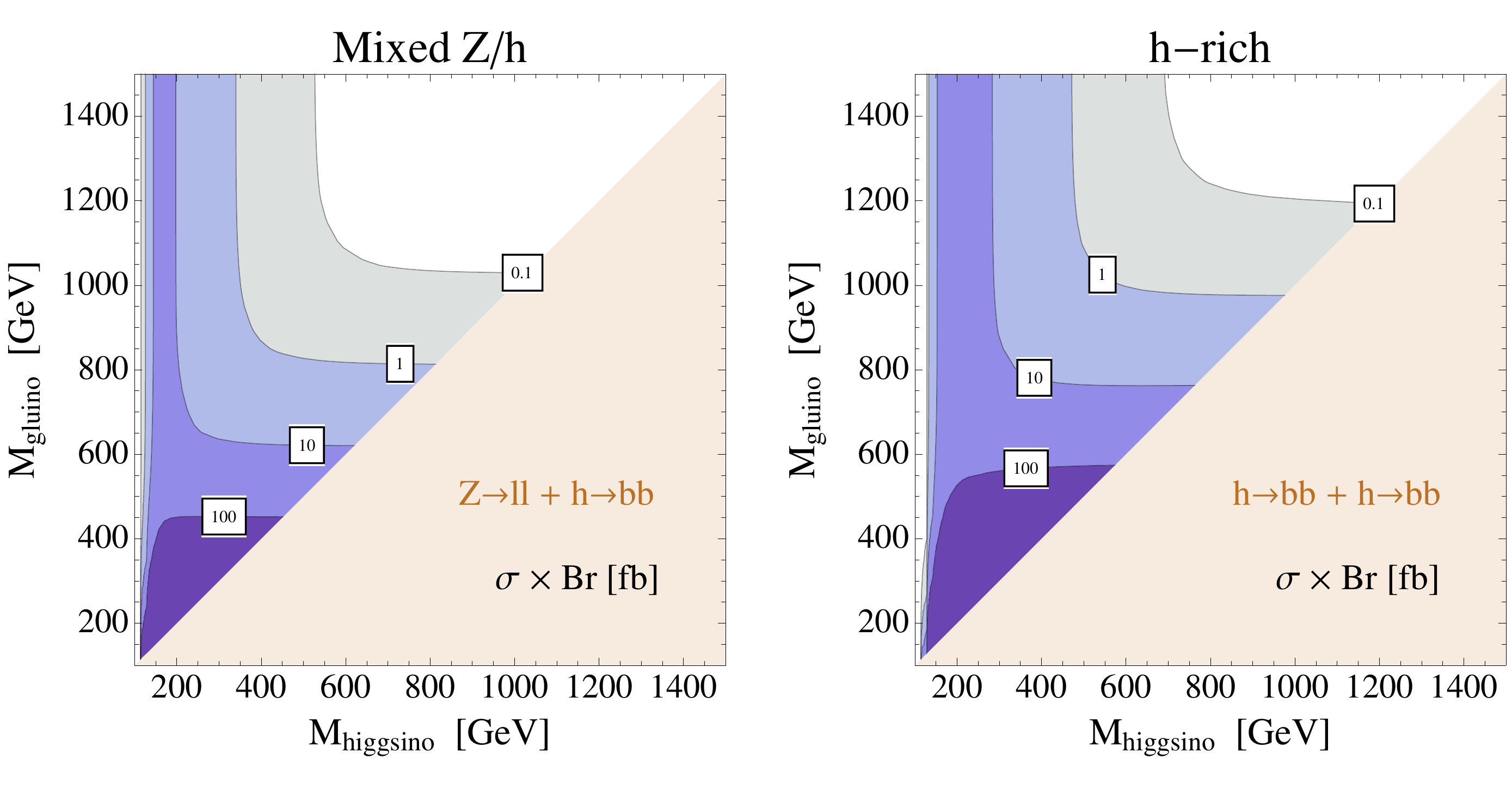}
\end{center}
\caption{$\sigma \times \mathrm{Br}$ into the $Z(\ell^+\ell^-) + h(b\bar b)$ ({\it left}) and $h(b\bar b) + h(b\bar b)$ ({\it right}) final states, for the mixed $Z/h$ and $h$-rich higgsino benchmarks, described in the text.  These are promising final states to discover higgsino NLSPs that decay to higgses.}
\label{fig:higgsinoCross}
\end{figure}

Above, we focused on higgsinos that dominantly decay to $Z$ bosons.  In this section, we briefly discuss several more general possibilities, which highlight additional discovery channels.  Recall from above that the higgsino dominantly decays to $Z$'s at low $\tan \beta$ and positive $\mu$.  For larger values of $\tan \beta$, the higgsino decays to a roughly even mixture of $Z$'s and $h$'s.  For this mixed $Z/h$ scenario, a promising discovery mode is to search for a leptonic $Z$ plus (up to) two $b$-jets from the higgs, plus missing energy.  On the left of figure~\ref{fig:higgsinoCross}, we show the  $\sigma \times \mathrm{Br}$ for this final state (all cross-sections are NLO in this section), again for a simplified spectrum with a gluino and a higgsino, now taking $\tan \beta = 20$, $\mu < 0$. For the higgs sector we again take $m_h=120$ and decouple the other MSSM higgses.

Another interesting possibility is where the higgsino dominantly decays to higgses.  This occurs at low $\tan \beta$ and negative $\mu$.  A promising final state is double higgs production, $h \rightarrow b \bar b + h \rightarrow b \bar b$ plus missing energy, where up to four b-tags can be requested.   To the right of figure~\ref{fig:higgsinoCross}, we show the $\sigma \times \mathrm{Br}$ for this final state, setting $\tan \beta = 2$, $\mu < 0$, and again taking the decoupling regime for the higgs.  

\begin{figure}[b!]
\begin{center}
\includegraphics[scale=0.55]{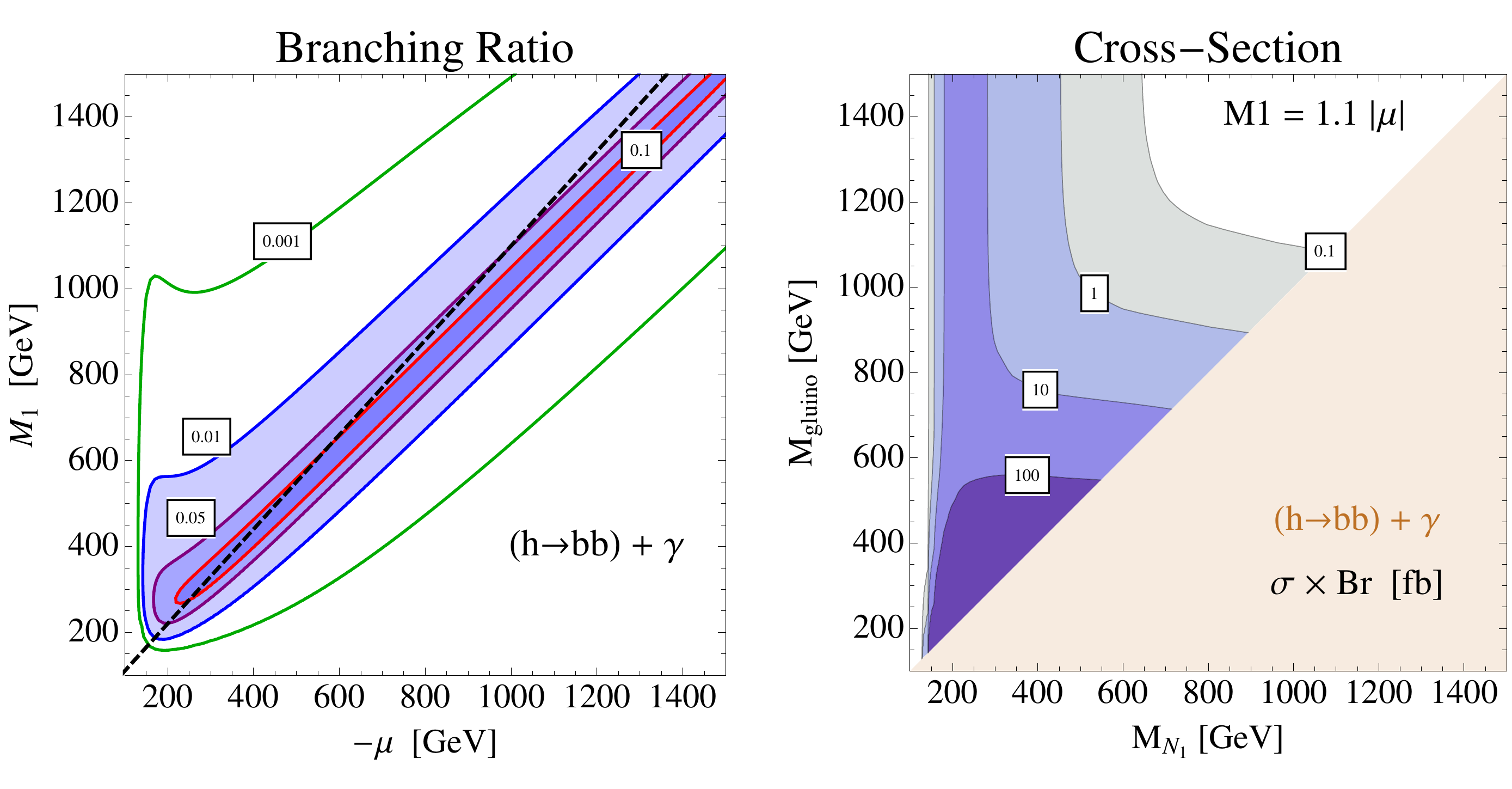}
\end{center}
\caption{The $h \rightarrow bb + \gamma$ final state can be used to discover an NLSP that is a higgsino/bino admixture.  The {\it left} plot shows the branching ratio into this final state as a function of $M_1$ and $\mu$, and we see that the branching ratio is only large when these parameters are comparable, $M_1 \simeq \mu$.  The {\it right} plot shows $\sigma \times \mathrm{Br}$ for our higgsino/bino admixture benchmark, described in the text.  In this benchmark, we fix $M_1 = -1.1 \times \mu$ with negative $\mu$, which is shown as a dashed black line on the left plot.}
\label{fig:HiggsinoBinoMix}
\end{figure}

One exciting scenario is that the LHC will discover an excess in one of the above final states, by relying on $b$-tags to suppress backgrounds.  Then, with more data, it may be possible to reconstruct the higgs with $b$-jet masses.  Therefore, if the NLSP is a  mixed $Z/h$ or higgs-rich higgsino, the discovery of supersymmetry may also include an early discovery of the higgs boson!  We also comment that if the higgs is heavy enough, $m_h \gtrsim 150$~GeV, there can be an appreciable branching ratio to $h \rightarrow W W^*$, and the $b's$ in the above topologies can be replaced with multilepton final states.

So far in this paper, we have focused on pure gauge eigenstate NLSPs, for simplicity.  But there are additional final state channels that can turn on when the NLSP is a mixture.  For example, if the NLSP is a bino/higgsino admixture, there can be a large branching ratio of the NLSP to both photons and higgses.  An NLSP of this type may be discovered in the $\gamma + 2 b$ plus missing energy final state.  To the left of figure~\ref{fig:HiggsinoBinoMix}, we show the branching ratio into this final state, as a function of $M_1$ and $\mu$, fixing $\tan \beta = 20$ and taking $\mu < 0$.  We see that a large branching ratio requires some tuning, $M_1 \approx | \mu | $.  To the right of figure~\ref{fig:HiggsinoBinoMix}, we show $\sigma \times \mathrm{Br}$ into this final state, fixing $M_1 = -1.1 \mu$.

\section*{Acknowledgments} 
We gratefully acknowledge Spencer Chang, Yuri Gershtein, Eva Halkiadakis, Mariangela Lisanti, Jason Nielsen, Natalia Panikashvili, Michael Park, Matt Reece, Bruce Schumm, Tracy Slatyer, Scott Thomas, Chris Tully,  Kai Wang, and Yue Zhao for helpful conversations. We especially thank Yuri and Eva for useful conversations, and for carrying out the $\ell+\gamma+\met$ search at CMS\@!  The research of DS was supported in part by a DOE Early Career Award.

\appendix
\section{Comments on Non-minimal Spectra}
\label{app:nonminspec}

In this paper, we have considered a set of minimal spectra for general neutralino NLSPs in gauge mediation. We have used these minimal spectra as a sandbox to explore inclusive searches for general GMSB signatures. However, it is also worth considering what can happen with non-minimal spectra. In this appendix, we will comment briefly about the differences in using non-minimal spectra, focusing for simplicity on the wino co-NLSP scenario. 

\begin{figure}[t!]
\begin{center}
\includegraphics[scale=0.5]{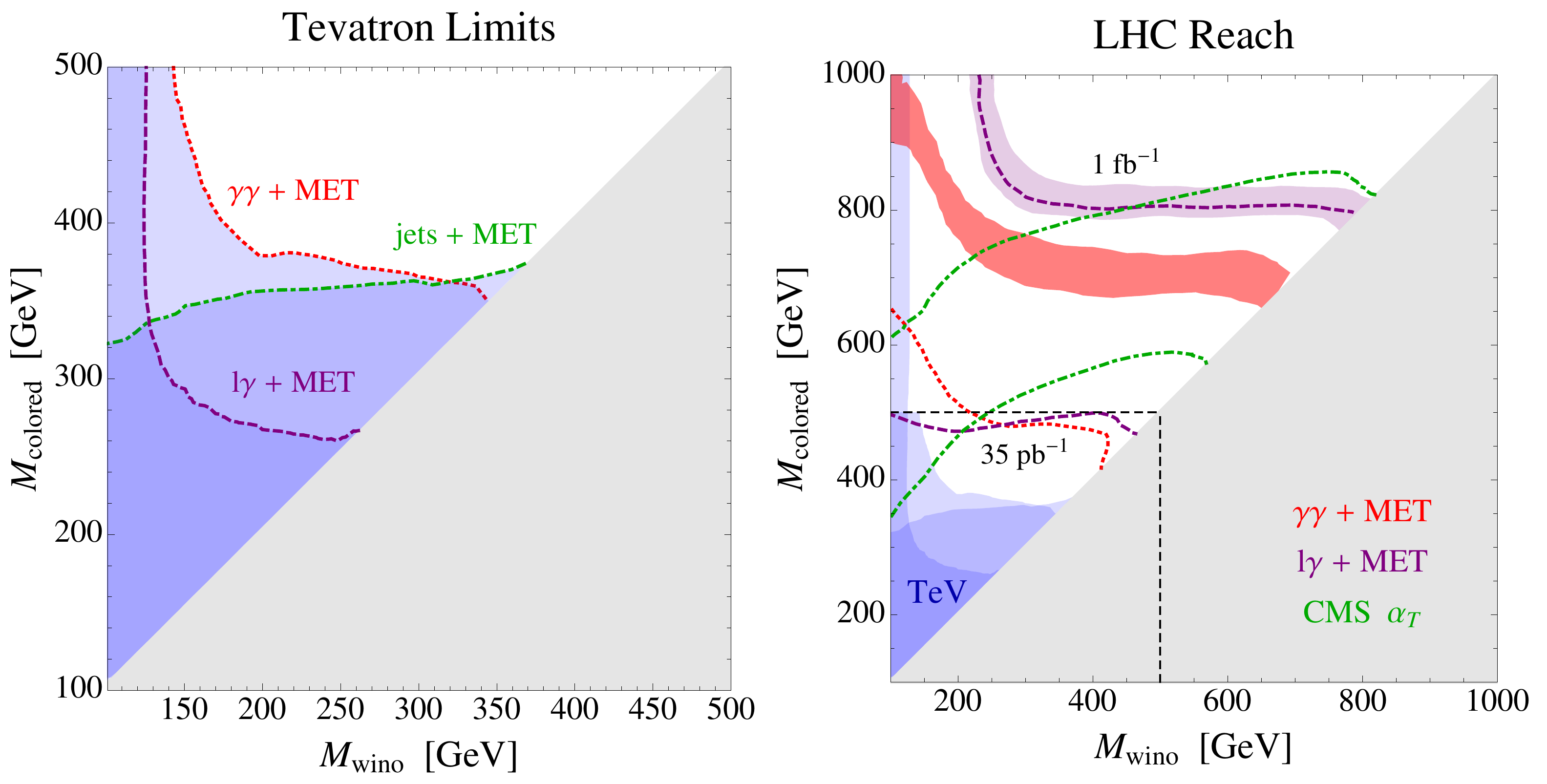}
\end{center}
\caption{The Tevatron limit ({\it left}) and LHC reach ({\it right}), for a non-minimal wino co-NLSP benchmark where the squarks are also light.}
\label{fig:winoreachmQ}
\end{figure}

The first major difference that can occur when using non-minimal spectra is different rates for Tevatron vs.\ LHC\@. In the body of the paper, we decoupled all the colored states except the gluino, for simplicity. However, bringing down the squarks can greatly enhance the production cross-section of the LHC over the Tevatron. This in turn can lead to more optimistic projections and stronger motivations for early LHC searches. Shown in figure~\ref{fig:winoreachmQ} are the Tevatron limits and LHC reaches in a non-minimal scenario, in which the squarks are also light. Specifically, we have decoupled the up-type squarks in order to satisfy the GGM sum rules~\cite{Meade:2008wd}; the remaining squark masses $m_{Q_L}$ and $m_{\bar d_R}$  are set to 25 GeV below the $m_{\tilde g}$ (the slight offset is to ensure that gluino decays are simple). We see from figures~\ref{fig:winoreachmQ} that while the Tevatron limits are similar to the decoupled squark case shown in figure~\ref{fig:winoreach}, the LHC reach contours are significantly enhanced. 

A second major difference in using non-minimal spectra was already mentioned in section~\ref{sec:WinoLHC}. Including different intermediate sparticles can affect the relative importance of the different final states. This is especially stark in the case of non-inclusive searches such as jets~+~$\met$, which typically veto on leptons (and also photons, in the case of the CMS $\alpha_T$ search) to suppress backgrounds from $W+{\rm jets}$. In our minimal spectrum the only source of leptons was from $\tilde W^\pm \to W^\pm+\tilde G$ decays. But if there exist intermediate sleptons in the spectrum, it is possible for most events to contain extra leptons, thereby greatly diminishing the sensitivity of jets~+~$\met$ type searches. This sensitivity to additional states should be kept in mind when attempting to compare the power of different final states.  Sometimes, the relative performance between two different searches, within a particular simplified  parameter space, can be misleading.


\end{document}